\documentclass[aps,prl,preprint,superscriptaddress]{revtex4-1}
\usepackage{graphicx}
\usepackage{dcolumn}
\usepackage{bm}
\usepackage{color}

\begin{document}

\title{Quantum statistics and   networks  by
asymmetric preferential attachment of
 nodes-between
bosons and fermions
}


\author{Masato Hisakado}
\email{hisakadom@yahoo.co.jp} 
\affiliation{
* Nomura Holdings, Inc., Otemachi 2-2-2, Chiyoda-ku, Tokyo 100-8130, Japan} 

\author{Shintaro Mori}
\email{shintaro.mori@gmail.com}
\affiliation{
\dag Department of Mathematics and Physics,
Graduate School of Science and Technology, 
Hirosaki University \\
Bunkyo-cho 3, Hirosaki, Aomori 036-8561, Japan}


\date{\today}

\begin{abstract}
In this article, we discuss the random graph, Barab\'{a}si-Albert (BA) model,
  and lattice networks from a unified view point,
  with the parameter $\omega$ with values $1,0, -1$ characterizing these networks,
  respectively.
The parameter is related to the preferential attachment of
 nodes in the networks and has different weights for the
 incoming and outgoing links.
 In addition, we discuss the correspondence between quantum statistics
 and the networks. 
 Positive and negative $\omega$ correspond to Bose and Fermi-like statistics,
 respectively, and we obtain the distribution that connects the two.
 When $\omega$ is positive, it is related to the threshold of Bose-Einstein
 condensation (BEC). As $\omega$ decreases, the area of the BEC phase is narrowed,
 and disappears in the limit $\omega=0$.
 When $\omega$ is negative, nodes have limits in the number of
 attachments for newly added nodes (outgoing links), which corresponds to
 Fermi statistics.
 We also observe the Fermi degeneracy of the network.
When $\omega=-1$, a standard Fermion-like network is observed.
Fermion networks are realized in the cryptocurrency network ``Tangle.’’ 
\hspace{0cm}
\vspace{1cm}

\end{abstract}


\maketitle


\subsection{I. Introduction}
Complex networks, especially networks with hubs, have been studied
extensively in the last two decades. The degree distribution in the networks
is important, because it is related with the features that define
  scale-free networks. Scale-free networks have hubs with a
  large number of links,
  and they are crucial in many applications.
  Complex networks can be applied in multidisciplinary areas such as sociology,
  social psychology, ethnology, and economics \cite{nw,MN}. 
  In statistical physics, complex systems can be described in terms of such
  networks. Studies on such topics have led to the development of
new fields of research such as sociophysics \cite{galam} and
econophysics \cite{Cont,Egu,Stau,Curty,nuno}, where financial markets
and opinion dynamics on networks are studied \cite{net, SF, DN}.
As a model for scale-free network, Barab\'{a}si and Albert (BA)
  model is well known \cite{BA}. It is an evolving network model and new nodes are
  linked with the network based on the preferential attachment process, where
  the probability that a node is selected by a new node is proportional
  to its degree. In addition, by introducing the fitness of each node
  to modify the probability, the relation with Bose statistics has been found \cite{BA1,BA2}.
  The relation between networks and Fermi statistics has also beed studied \cite{Bi,Ci,Hal}.

Herein, we discuss several networks from a unified perspective.
We propose a model that connects the random graph, Barab\'{a}si-Albert (BA) model,
 and lattice model with a parameter $\omega$, which represents the weights of
  the incoming and outgoing links. Our model is an evolving network model,
   and newly added nodes are more likely to be linked to other nodes
   with high ``popularity.’’. The popularity is defined as
   the sum of the number of incoming and the product of the number of
   outgoing links with $\omega$.
 When $\omega$ is positive, the evolving mechanism is the preferential attachment
process in the BA model, and the model exhibits positive feedback.
When $\omega=1$, the weights for the incoming and outgoing
links are equal, and the model takes the form of the BA model \cite{BA}.
When $\omega$ is negative, there is a limit on the number of the
  links from newly added nodes in the evolution of the network. 
In this case, the model incorporates negative feedback. In addition, $\omega$ is
related to the length of the memory, and indicates the age of the node to which a new node connects.
Long and short memories indicate that new nodes tend to connect to older and newer nodes, respectively.
When $\omega$ is positive, the network has long memory, and this corresponds to scale-free networks.
Networks with negative $\omega$ have intermediate memories, and the limit $\omega=-1$ corresponds
to extended lattices.

To understand $\omega$ more deeply, we introduce the fitness model, which is an extension of
the BA model\cite{BA1,BA2}.
Fitness is introduced in each node to adjust popularity, and
it can be transformed to the ``energy’’ using the Boltzmann weight. 
The fitness decreases as the energy increases, and the degree distribution can thus be
  interpreted in terms of quantum statistics. When $\omega$ is positive, the number of
  outgoing links are not limited, as in the case of Bose statistics. When $\omega$ is negative,
the number of outgoing links are limited, as in the case of Fermi statistics,
  where an energy level can accommodate one particle at most.
We obtain a distribution that connects the Bose and Fermi statistics.
When $\omega$ is negative, Fermi degeneracy can be observed in the network at
low temperatures. Nodes with high fitness have the maximum number of links, and nodes
with low fitness have no links except for the incoming links. When $\omega$ is
positive, Bose-Einstein condensation (BEC) was realized in the network, and
a small number of hubs with low energy (high popularity) is connected to most of the nodes in the network.
In other words, the ``winner takes all’’ scenario occurs.
Furthermore, $\omega$ is related to the threshold of BEC, and the BEC phase becomes
wider as $\omega$ increases. When $\omega=0$, which corresponds to the random network,
we obtain the Boltzmann distribution.

The remainder  of this paper is organized  as follows.
In section 2, we introduce  the networks which have the parameter $\omega$.
In the section 3,   we discuss the relation between the quantum statistics and networks.
Finally, the conclusions are presented in section 4.

\subsection{II. Networks}

\begin{figure}[h]
\includegraphics[width=110mm]{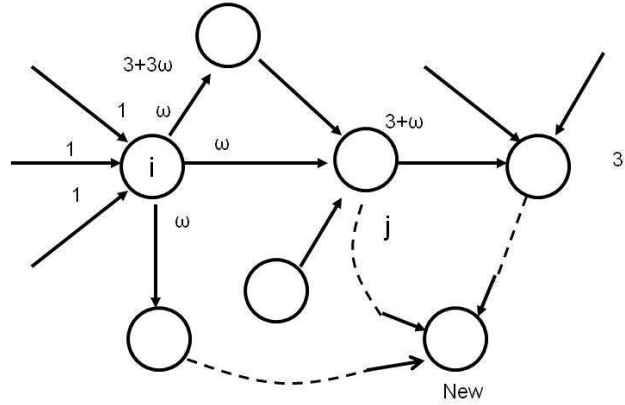}
\caption{Sample graph with $r=3$. Node $i$ selects three nodes and
  is selected by three nodes. The arrows represent the directions
  from the selected nodes to the selecting nodes.
  $k_{i}^{IN}=k_{i}^{OUT}=3$, and $l_i=3\omega+3$.
Node $j$ selects three nodes and is selected by one node.
$k_{j}^{IN}=3,k_{j}^{OUT}=1$, and $l_j=3+\omega $. $l_j$
changes from $3+\omega$ to $3+2\omega$ when a new node
  joins to the network and selects node $j$.
  $k_{j}^{OUT}$ increases from $1$ to $2$.}
\label{BA2}
\end{figure}

We begin this section by discussing the generation of the networks.
The proposed model is an evolving network model, and we
consider the case where a newly added node selects at most
$r$ different nodes based on the popularity. The process is sequential
like in the BA model \cite{BA, BA1.5}.
When the $i$-th node enter the network
at time $t=i$, it selects $r$
nodes for the connections, as shown in Fig\ref{BA2}.
For node $i$, the links from the selected $r$ nodes are
incoming links,
and these are shown by the incoming arrows in Fig.\ref{BA2}.
For the $r$ nodes selected by node $i$, the links to the node $i$
are outgoing links.
In Fig.\ref{BA2}, we draw the links with the outgoing arrows.
We denote the number of incoming and outgoing links of
node $i$ as $k_{i}^{IN}$ and $k_{i}^{OUT}$, respectively.
The degree of node $i$ is given by $k_{i}=k_{i}^{IN}+k_{i}^{OUT}$.
The popularity $l_i$ of node $i$ is defined as
\begin{equation}
l_{i}=\bar{\omega}\cdot k_{i}^{IN}+\omega \cdot k_{i}^{OUT}. \label{eq:pop}
\end{equation}
Here, $\bar{\omega}$ and $\omega$ are the weights for $k_{i}^{IN}$
and $k_{i}^{OUT}$, respectively. When $\omega=\bar{\omega}=1$,
the model attains the form of the BA model. Hereafter,
we set $\bar{\omega}=1$ for normalization without loss of
generality.

The initial state of the model is the complete network with $r+1$ nodes,
and we label them as node $i,i=1,\cdots,r+1$.
We set $k_{i}^{IN}=i-1$ and $k_{i}^{OUT}=r+1-i$.
We add node $t\ge r+2$ and link it with $r$ nodes sequentially.
The probability that node $i<t$ is selected by node $t$ is
\begin{equation}
P(\mbox{node} \,\, i \,\, \mbox{is selected by node}\,\,t)
=p_{i}(t)=\frac{\mbox{Max}(l_i,0)}{\sum_{s=1}^{t-1}\mbox{Max}(l_s,0)} \label{eq:prob}
\end{equation}
We note that when $l_i$ becomes negative or zero,
node $i$ cannot be selected. This occurs for negative values of
$\omega$ and $k_{i}^{OUT}\ge \frac{k_{i}^{IN}}{|\omega|}$.
The maximum value of $k_{i}^{OUT}$ is $\lceil k_{i}^{IN}/|\omega| \rceil$,
where $\lceil x\rceil$ is the ceiling function, and $k_{i}^{OUT}$ cannot
exceed $\lceil \frac{k_{i}^{IN}}{|\omega|} \rceil$. We denote the number
of the nodes with positive $p_{i}(t),i=1,\cdots,t-1$ by $r'$. If $r'$
is less than or equal to $r$, all nodes are linked by node $t$, and we set
$k_{t}^{IN}=r'$. If $r'>r$, $r$ different nodes are selected
from $r'$ nodes with probability $p_{i}(t)$, and we set $k_{t}^{IN}=r$.
$k_{i}^{OUT}$ for the selected nodes increases by unity.

$\omega=1$ and $\omega=0$ correspond to the BA model \cite{BA} and
the random network, respectively. This is because when
$\omega=0$, all nodes are selected with equal probability.
We consider that the range of negative $\omega$ is $-1\leq\omega<0$.
When $\omega>0$, the popularity and the probability of
selection increase during node selection.
This is called positive feedback or preferential attachment.
When $\omega<0$, the popularity and the probability selection
decrease during node selection.

\begin{figure}[h]
\begin{center}
\begin{tabular}{c}
\begin{minipage}{0.5\hsize}
\begin{center}
\includegraphics[clip, width=8cm]{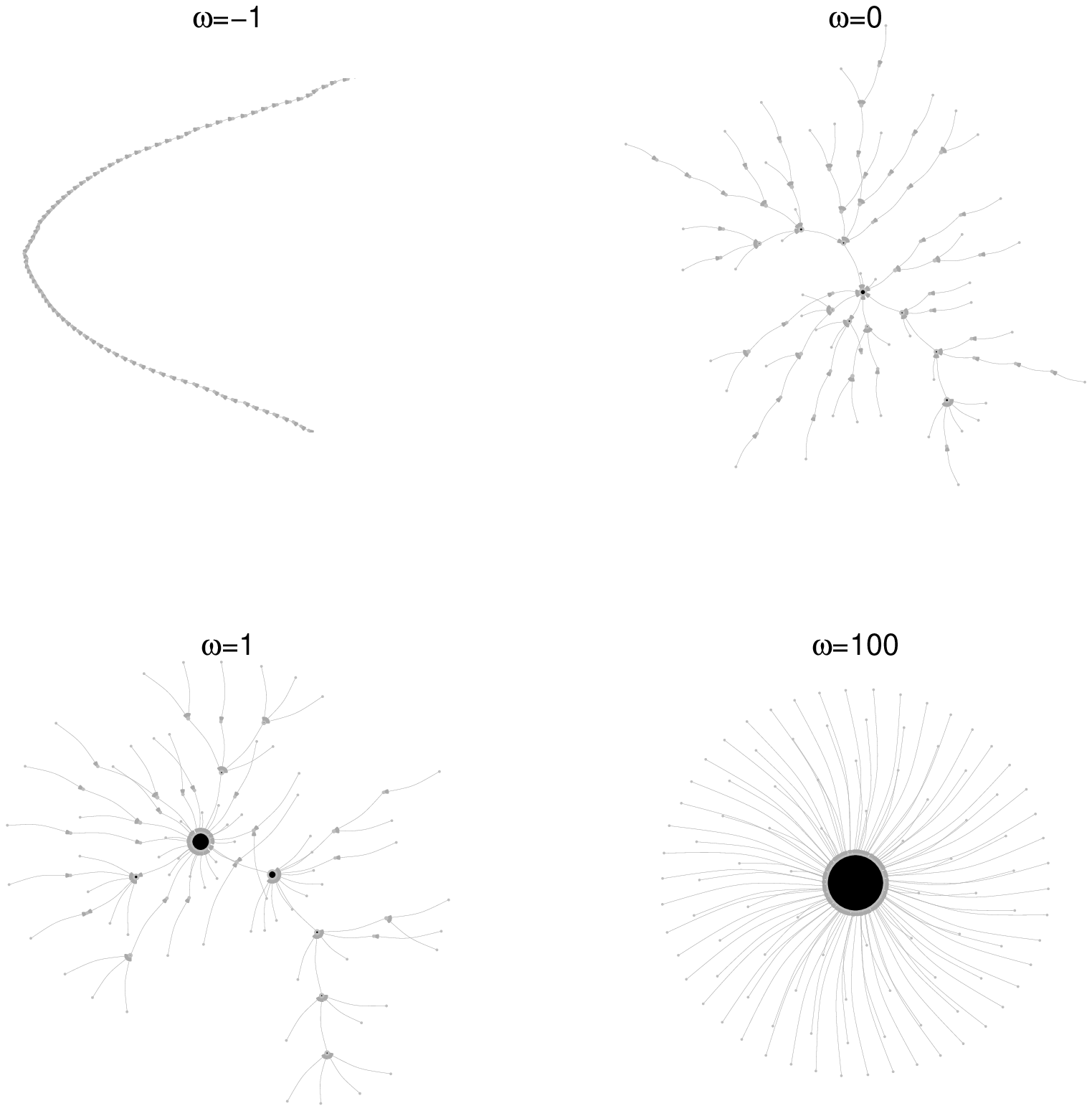}
\hspace{1.6cm} (a)
\end{center}
\end{minipage}
\begin{minipage}{0.5\hsize}
\begin{center}
\includegraphics[clip, width=8cm]{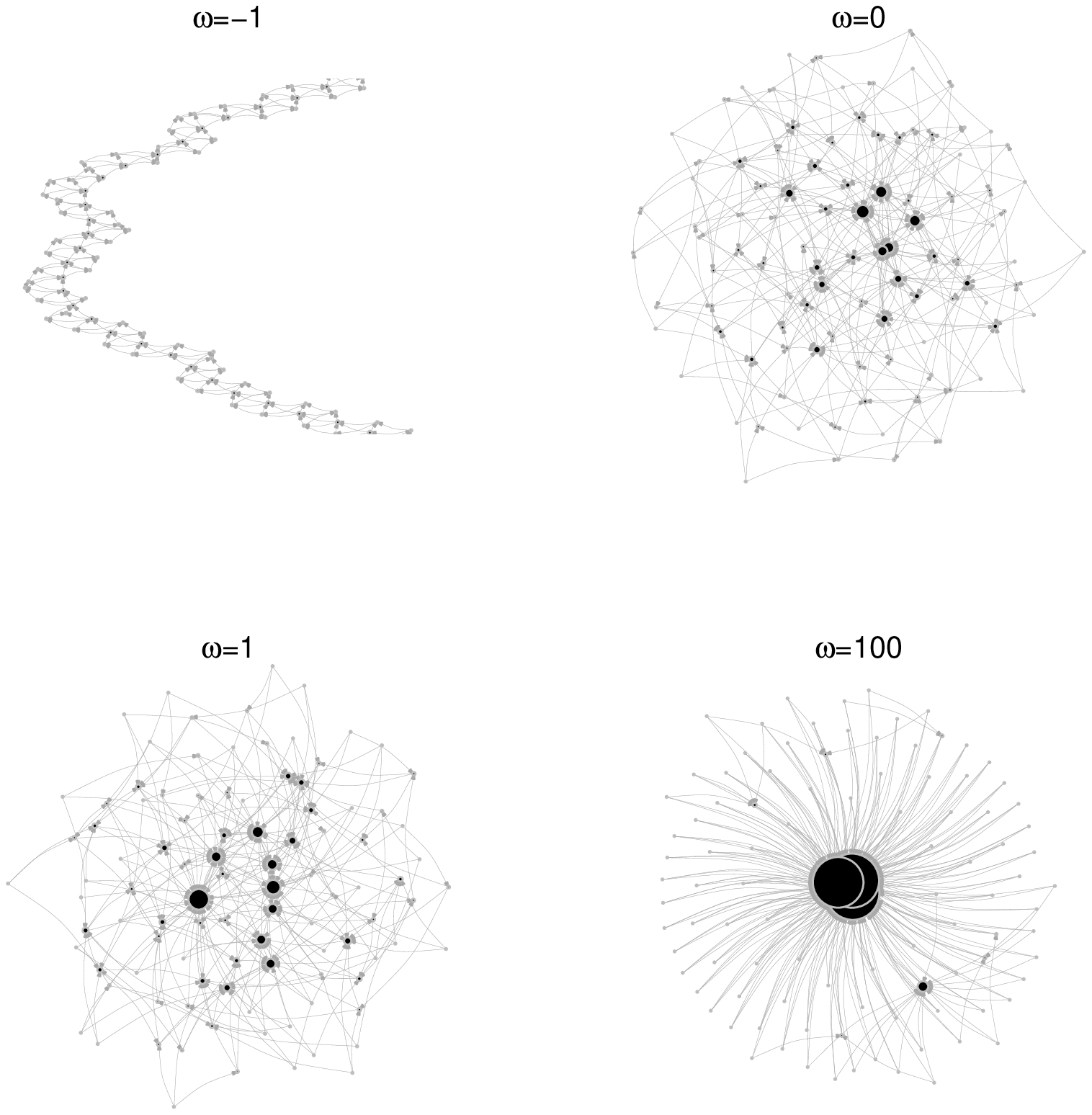}
 \hspace{1.6cm} (b)
\end{center}
\end{minipage}
 \end{tabular}
\caption{Sample networks with $\omega \in \{-1,0,1,100\}$ in 100 steps.
  (a) $r=1$. In this case, the network becomes a tree network. (b) $r=3$.
  $\omega=1$ corresponds to the BA model. When $\omega>0$, the network is a scale-free network.
  When $\omega=0$, the network is a random network.
  When $\omega=-1$, the network corresponds to an extended lattice.}
\label{sample}
\end{center}
\end{figure}

Samples of the network are presented in Fig.\ref{sample}.
In Fig.\ref{sample} (a), we show the case of $r=1$,
which corresponds to tree networks, and in (b), we show the case of $r=3$.
We upload the R script that perform the numerical studies of
this paper \cite{git}.

\subsection{A. Degree distribution} 
We consider the degree distribution of the networks. The number of links in the
node $i$ is $k_i$. When $\omega=1$ for the BA model, $l_i=k_i$. The sum of the
popularities is $\sum_j l_j=(\omega+1)r[t-1/2(r+1)]\sim (\omega+1)rt$ for large $t$.
Note that when $\omega$ is negative, the popularity decreases as the number of links to
  the node increases. However, the total popularity does not decrease when $\omega>-1$,
  and new nodes can join. We can obtain the differential equation for the expected
value of $k_i$ as
\begin{equation}
\frac{d k_i}{dt}=\frac{(k_i-r)\omega+r}{(1+\omega)t}.
\label{df}
\end{equation}

Note that when $\omega$ is negative, if $k_{i}^{OUT}>k_{i}^{IN}/|\omega|$,
$l_i=k_{i}^{IN}-|\omega| k_{i}^{OUT}<0$ and $p_{i}=0$.
We discuss this effect in more detail later.

\subsubsection{i. Case when $\omega>0$ }
\begin{figure}[htbp]
\begin{center}  
  \begin{tabular}{cc}
      \begin{minipage}{0.5\hsize}
        \begin{center}
          \includegraphics[clip, width=6cm]{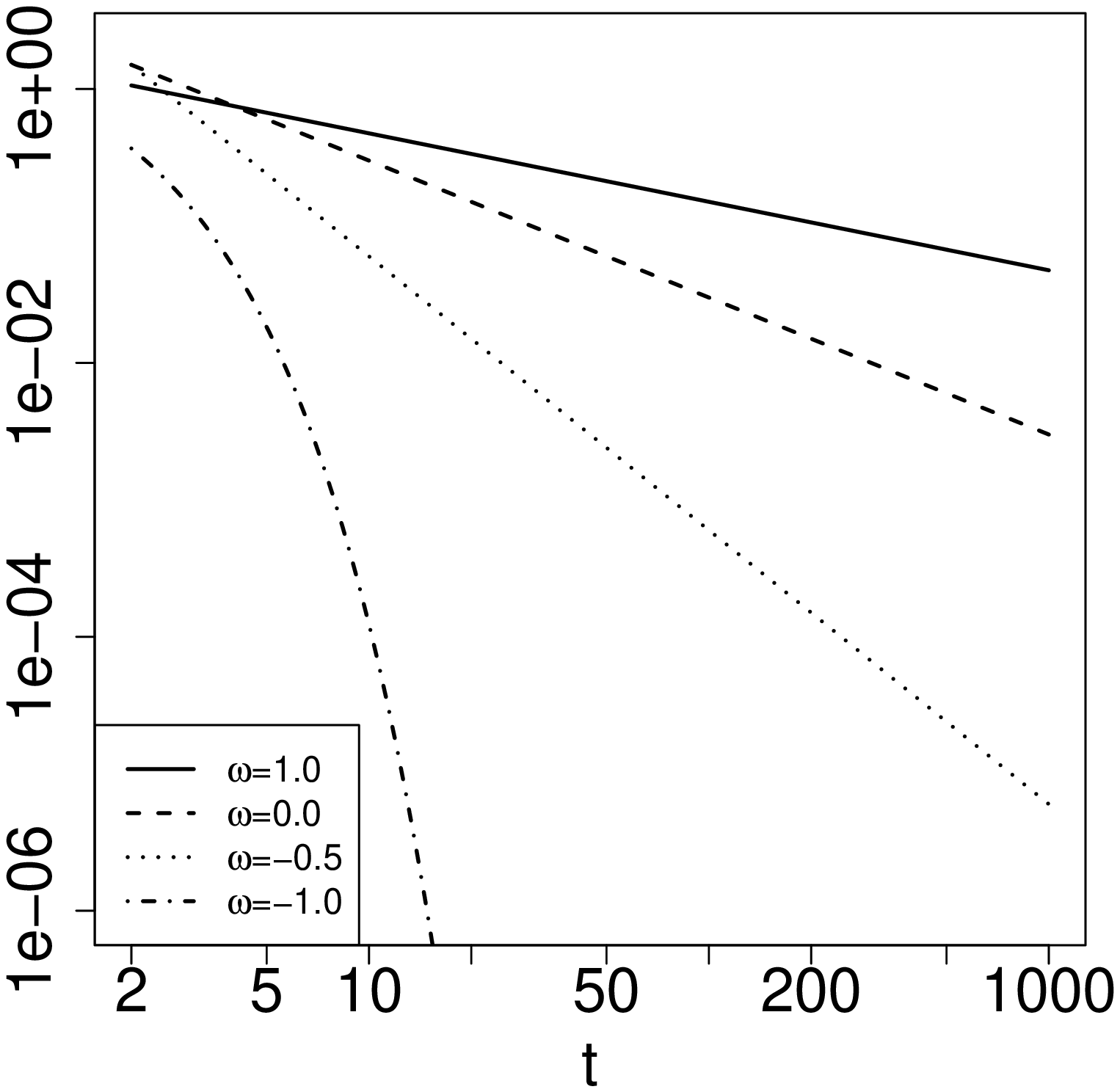}
          \hspace{-3cm}(a)
        \end{center}
      \end{minipage}
      &
      \begin{minipage}{0.5\hsize}
        \begin{center}
          \includegraphics[clip, width=6cm]{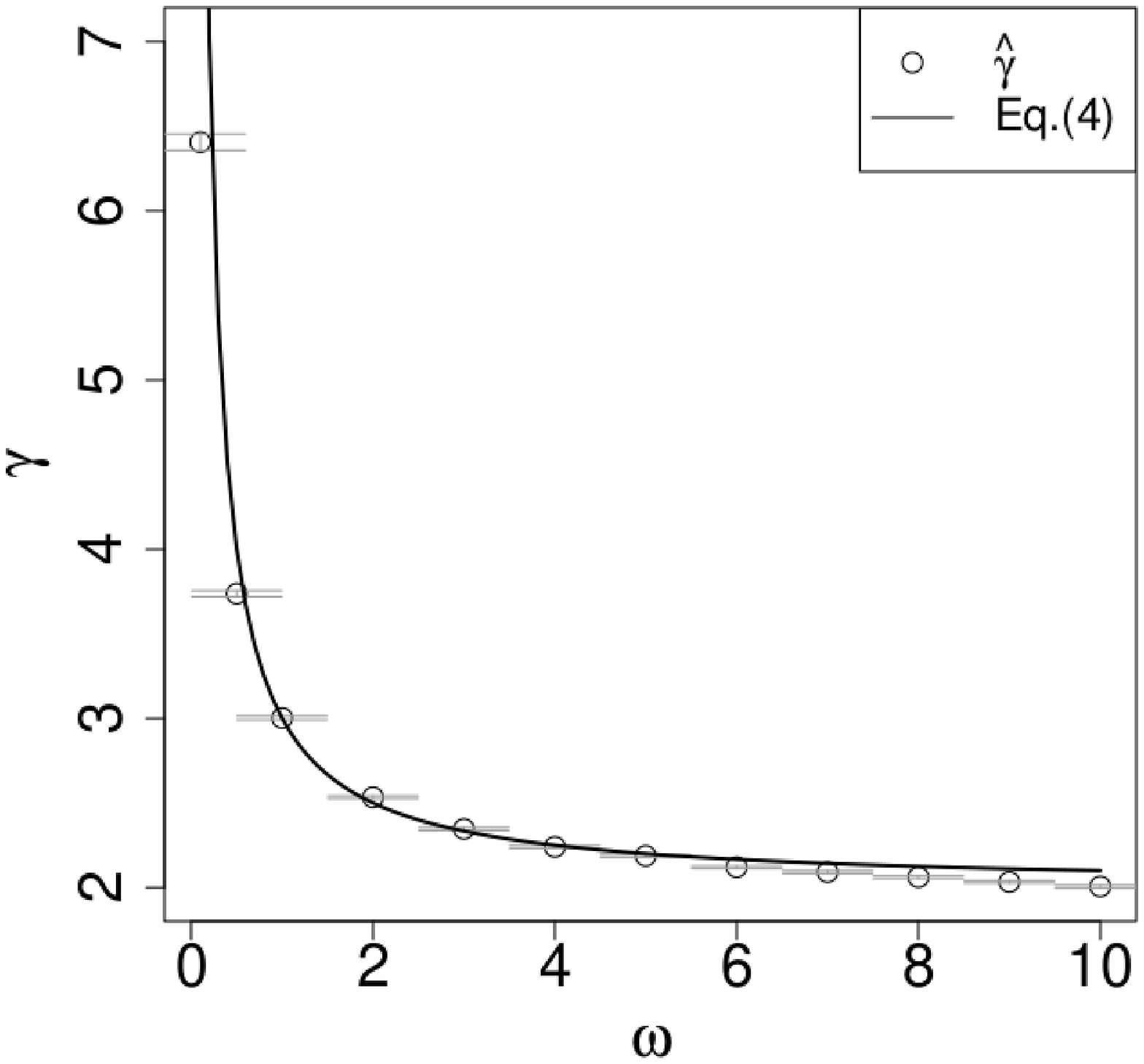}
          \hspace{-3cm}(b)
        \end{center}
      \end{minipage}
      \\
      \begin{minipage}{0.5\hsize}
        \begin{center}
          \includegraphics[clip, width=6cm]{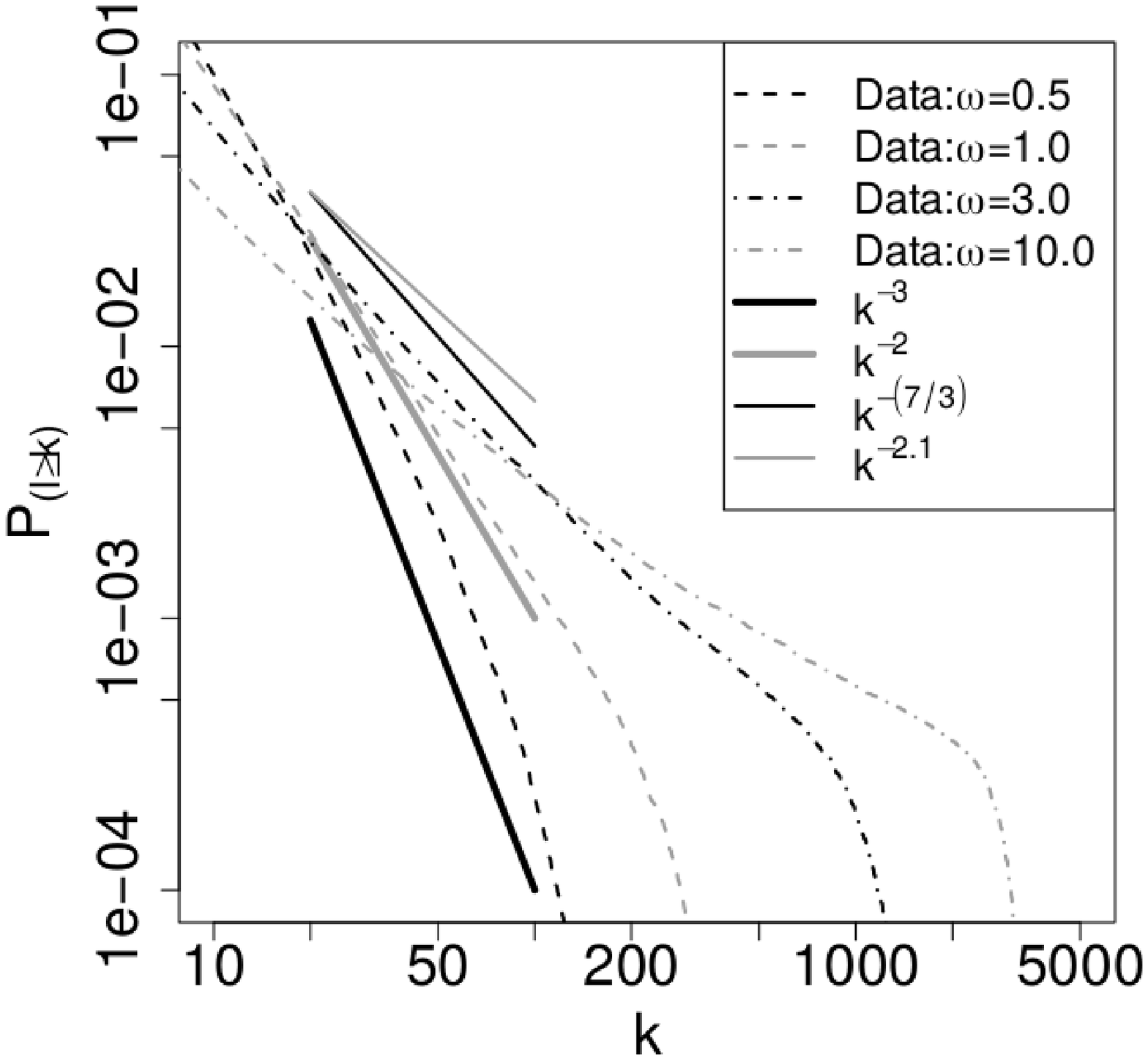}
          \hspace{-3cm}(c)
        \end{center}
      \end{minipage}
      &
      \begin{minipage}{0.5\hsize}
        \begin{center}
          \includegraphics[clip, width=6cm]{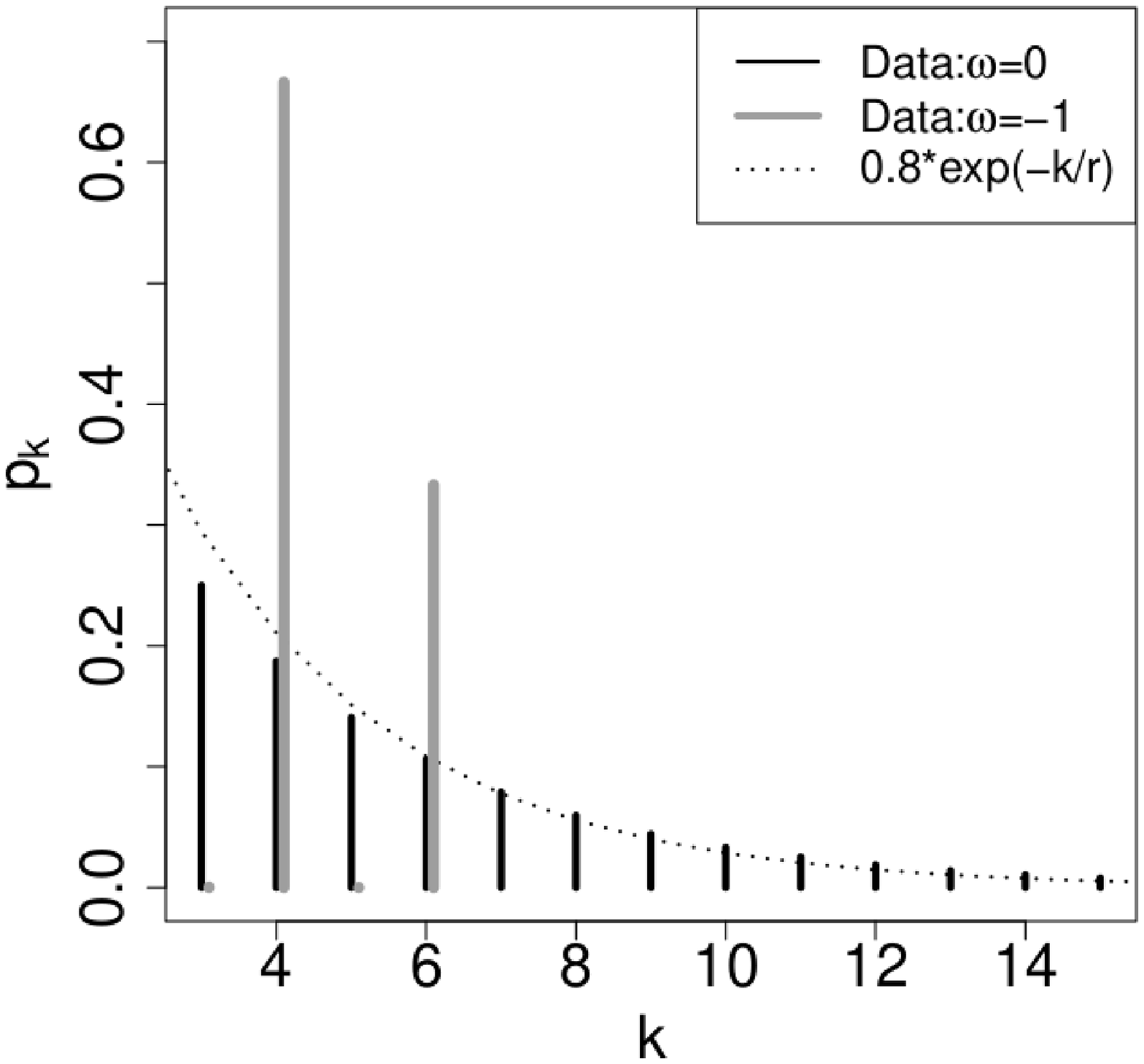}
          \hspace{-3cm}(d)
        \end{center}
      \end{minipage}
\end{tabular}
  \caption{(a)Plot of the average probability of the choice of node $i$ (memory) vs. $t$.
    The power index of the decay of the memory is $1/(1+\omega)$.
(b) Plot of $\gamma$ vs. $\omega>0$, which is the power index of the degree distribution in Eq.(4).
    We show the estimates of $\gamma$ and its standard errors using symbols ($\circ$) and error-bars. We sample 100 networks of $10^4$ nodes for each $\omega\in \{0.1,0.5,1,2,3,4,5,10\}$,
      and estimate $\gamma$ using the empirical degree distributions.
      (c) Plot of the empirical degree distributions for $\omega=0.5,1,3,10$ for the data in (b). The straight lines are the power-law equations with the theoretical power index $\gamma=2+1/\omega$.
(d) Plot of the empirical distribution for $\omega=0,-1$.
}
 
\label{memory}
\end{center}
\end{figure}

Eq.(\ref{df}) can be solved with the initial condition $k_i(t_i)=r$ and we
obtain,
\[
k_i=\frac{r}{\omega}\left(\frac{t}{t_i}\right)^{\omega/(1+\omega)}-r\frac{1-\omega}{\omega}.
\]
We estimate the average probability that node $i$ is selected as
\[
\frac{l_i}{(1+\omega)t}=\frac{r}{1+\omega}\left(\frac{1}{t_i}\right)^{\frac{\omega}{1+\omega}}t^{-1/(1+\omega)}.
\]
The average frequency of selection of node $i$,  called ``memory,’’
decreases as a power law with $t$, and the index is $1/(1+\omega)$.
We summarize the power index of the memory $1/(1+\omega)$ in Fig.\ref{memory}(a).
As $\omega$ increases, the power index becomes smaller, and the memory becomes longer.
In the limit $\omega\to \infty$, the memory covers all history equally,
and the power index of the memory becomes 0.
We term the cases where the power index is larger than 1
and smaller than 1 as the cases of intermediate memory and
long memory, respectively. 
This classification depends on whether the integral of the memory over $t$
is finite\cite{Long,Hisakado6,Hisakado7}.

We can obtain the cumulative degree distribution of the network as

\begin{eqnarray}
  P[k_i(t)<k]&=&P\left[(\frac{r}{\omega})^{(1+\omega)/\omega}
    \frac{t}{(k+r\frac{1+\omega}{\omega})^{(1+\omega)/\omega}}<t_i\right]
\nonumber \\
&=&
1-(\frac{r}{\omega})^{(1+\omega)/\omega}
\frac{1}{(k+r\frac{1-\omega}{\omega})^{(1+\omega)/\omega}}\nonumber.
\end{eqnarray}

Note that $k_i\geq r$.
Hence, the degree distribution is
\begin{equation}
p(k)=\frac{\partial P[k_i(t)<k]}{\partial k} 
\propto k^{-\frac{1+2\omega}{\omega}}.
\end{equation}
In this region, the degree distribution exhibits power-law decay.
When $\omega=1$, the network corresponds to the BA model with distribution $k^{-3}$.
When $\omega\rightarrow\infty$ and $\omega=1/2$, the networks
have distributions $k^{-2}$ and $k^{-4}$, respectively.
We show the power index in Fig.\ref{memory} (b).
We also show the numerical estimates of $\gamma$ with 100 sample networks
of $10^4$ nodes with $r=3$. We estimate $\gamma$ and its standard error
using the maximum likelihood estimator\cite{newman}. This is similar
to the extended BA model \cite{beta,Krap,Krap2, Krap3,Doro,Doro2,Basu},
which has a discount factor for each node. If the discount factor is introduced,
  the transition from the lattice to the scale-free network can be observed.
  However, these networks do not include trees, whereas our networks include the tree network.
The minimum power index is 2 when the network is graphical. This means that a degree sequence
can be converted into a simple graph\cite{nw}. Fig.\ref{memory} (c) shows the empirical
  degree distribution for the data in Fig.\ref{memory} (b) with $\omega=0.5,1,3,10$.
  When $\omega=0.5$ or $10$, the discrepancy from the power-law behavior becomes apparent.
  This may be attributed to the finite size effect. It becomes difficult to observe the
  power-law behavior for a small $\omega<<1$ and a large $\omega>>1$.

\subsubsection{ii. Case when $\omega=0$}
We solve Eq.(\ref{df}) with the initial condition $k_i(t_i)=r$ when $\omega=0$ to obtain

\[
k_i(t)=r \log(\frac{t}{t_i})+r.
\]

We estimate the probability that node $i$ is selected as
\[
\frac{l_i}{t}=\frac{r}{t}.
\]
The selected frequency decreases as a power law, and the index is $1$.
Therefore, the memory is long, but shorter than the case when $\omega>0$.
We can obtain the cumulative degree distribution of the network as

\[
P[k_i(t)<k]=P[t_i>t e^{-\frac{k}{r}}]=
1- e^{-\frac{k}{r}+1}.
\]

The degree distribution is
\begin{equation}
p(k)\propto    e^{-\frac{k}{r}}.
\end{equation}
The degree distribution decays exponentially (Fig.\ref{memory}(d)), and 
the model corresponds to a random network\cite{nw}.

\subsubsection{iii. Case when $-1<\omega<0$}

In this case, there is a probability that the popularity $l_i$ becomes negative and $p_i=0$.
Eq.(\ref{df}) becomes an approximated equation when $1/|\omega|$  is not an integer.
When $1/|\omega|$ is an integer, $p_i\geq 0$ for all $i$.

We solve Eq.(\ref{df}) with the initial condition $k_i(t_i)=r$ for $\omega<0$ as
\[
k_i(t)= r+r\frac{1}{|\omega|}\left(1-
\left(\frac{t_i}{t}\right)^{\frac{|\omega|}{1-|\omega|}}\right).
\]
Note that in this case, $r\leq k_i  \leq r+\lceil r/|\omega| \rceil$.
We estimate the probability that node $i$ is selected as
\[
\frac{l_i}{(1+\omega)t}=\frac{r}{1+\omega}t_i^{-\frac{\omega}{1+\omega}}\cdot t^{-\frac{1}{1+\omega}}.
\]
The selected frequency of node $i$ decreases as per the power law with $t$, and the index is
$1/(1+\omega)$.
When $-1<\omega<0$, the power index is larger than 1, and the memory
becomes intermediate. We can obtain the cumulative degree distribution of the network as

\begin{eqnarray}
P[k_i(t)<k]&=&
P\left[(\frac{|\omega|}{r})^{\frac{1-|\omega|}{|\omega|}}
  (\frac{1+|\omega|}{|\omega|}r-k)^{\frac{1-|\omega|}{|\omega|}}t>t_i\right]
\nonumber \\
&=&
\left(1-\frac{|\omega|}{r}(k-r)\right)^{\frac{1-|\omega|}{|\omega|}}. 
\end{eqnarray}

\begin{figure}[h]
\includegraphics[width=120mm]{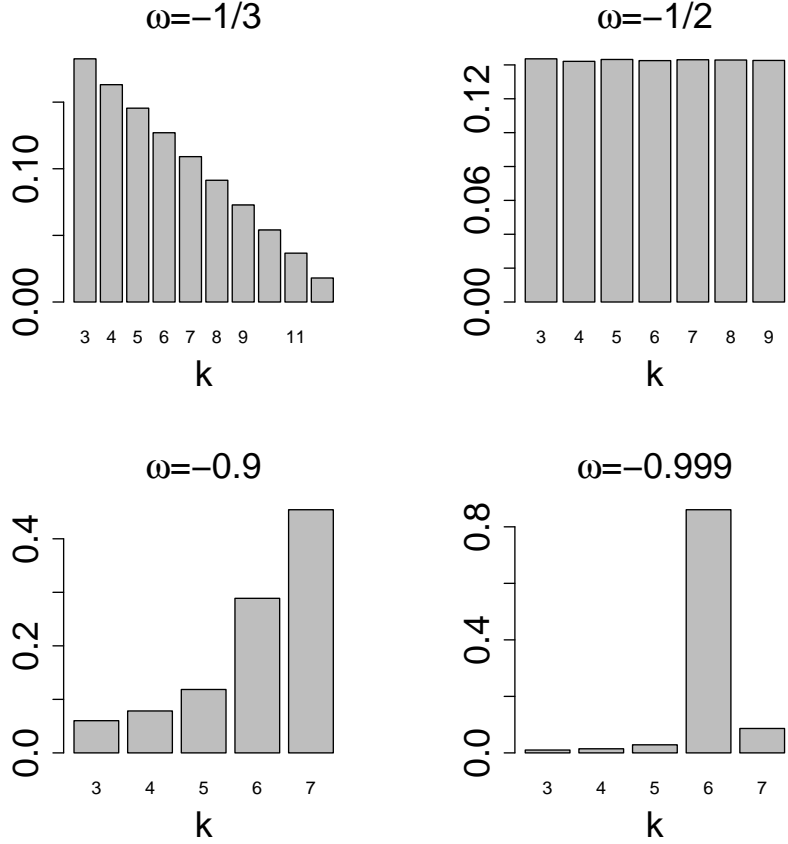}
\caption{Empirical degree distribution for $r=3$,
  $\omega=-1/3, -1/2, -0.9$, and $-0.999$. We sample $10^3$ networks of $10^4$ nodes.}
\label{degree}
\end{figure}

In this region, the distribution is the power decay in the range
$r$ to $r+\lceil r/|\omega| \rceil$.
The power index is $2-1/|\omega|$.
When $\omega=-1/2$, the power index is zero, and the distribution
becomes a uniform distribution between $r$ and $3r$.
We show the empirical degree distribution in Fig.\ref{degree} for $r=3$,
and confirm the uniform distribution for $\omega=-1/2$.
When $\omega=-0.9$ and $\omega=-0.999$, $k=6=3+\lceil 3/|\omega|\rceil$.
After the node $i$ is selected by four new nodes,
$l_i$ becomes negative, and $p_{i}=0$.
We can observe this effect to a larger extent for $\omega=-0.9$
than for $\omega=-0.999$. 

\subsubsection{iv. Limit $\omega\to -1$}

Before we discuss the limit $\omega\to -1$, we make a comment about the initial
configuration of the network.
If the initial condition of the network before its evolution is a complete graph,
the limit $\omega\to -1$ is trivial. The number of linkable nodes $r'$ is $r$ or $r-1$
and the network becomes an extended lattice. Here, we consider a more general
initial configuration of the network.
In the limit $\omega\to -1$, Eq.(\ref{df}) is undefined.
The differential equation for the expected value of $k_i$ becomes 
\begin{equation}
\frac{d k_i}{dt}=\frac{-(k_i-k_{i}^{IN})+k_{i}^{IN}}{C}=\frac{-(k_i-2k_{i}^{IN})}{C},
\label{df3}
\end{equation}
where $C$ is a constant that depends on the initial condition.
For example, $C$ is the number of nodes that have no links at time $0$.
We note that node $t$ cannot necessarily be linked with $r$ nodes when it
enters the network, as the number $r'$ of nodes with $p_i>0$ might be less than $r$.
The solution of Eq.(\ref{df3}) with $k(t_i)=k_{i}^{IN}$ is
\[
k_i=2k_{i}^{IN}-k_{i}^{IN} e^{-\frac{1}{C}(t-t_i)}.
\]
$k_i$ increases exponentially from $k_{i}^{IN}$ to $2k_{i}^{IN}$.
We estimate the probability that node $i$ is selected as
\[
\frac{l_i}{C}\sim \frac{r}{C}e^{-\frac{1}{C}(t-t_i)}.
\]
Here, we assume $k_{i}^{IN}=r$ for most nodes, which is true for $\omega>-1$.
The selected frequency decreases exponentially, and the memory becomes short.
We can confirm this in Fig.\ref{memory} (a).
In the case $C\rightarrow 0$, the network becomes an extended lattice,
which can be confirmed in Fig.\ref{memory} (d).
The degree distribution
is the superposition of delta functions at $k=2r$ and $2(r-1)$.
Most of the nodes are selected by the previous $r$ or $r-1$ nodes and
the numbers of the incoming and outgoing arrows are equal.

\subsection{B. Network properties}

\begin{figure}[h]
\begin{center}
\begin{tabular}{c}
\begin{minipage}{0.5\hsize}
\begin{center}
\includegraphics[clip, width=8cm]{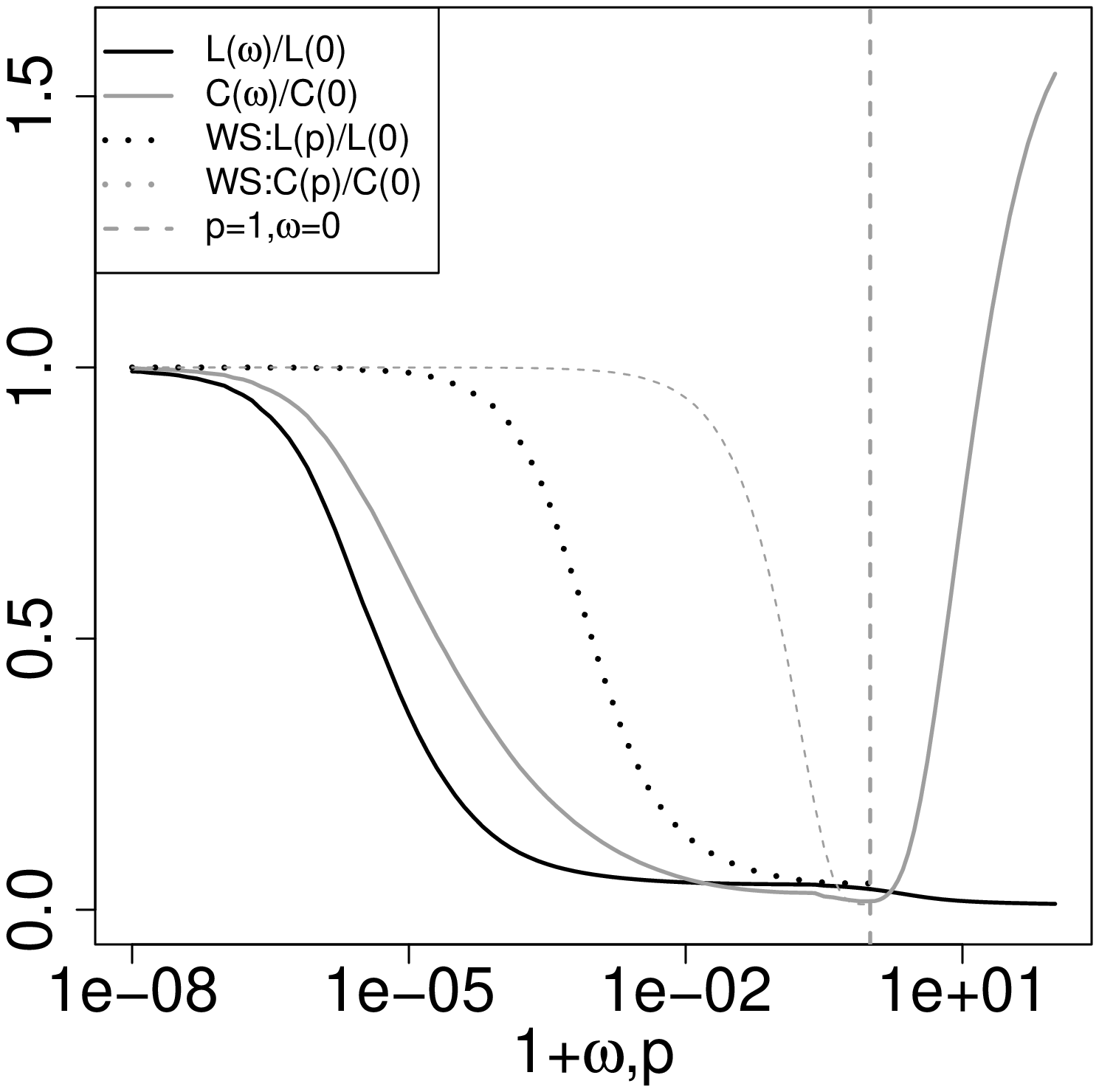}
\hspace{1.6cm} (a)
\end{center}
\end{minipage}
\begin{minipage}{0.5\hsize}
\begin{center}
\includegraphics[clip, width=8cm]{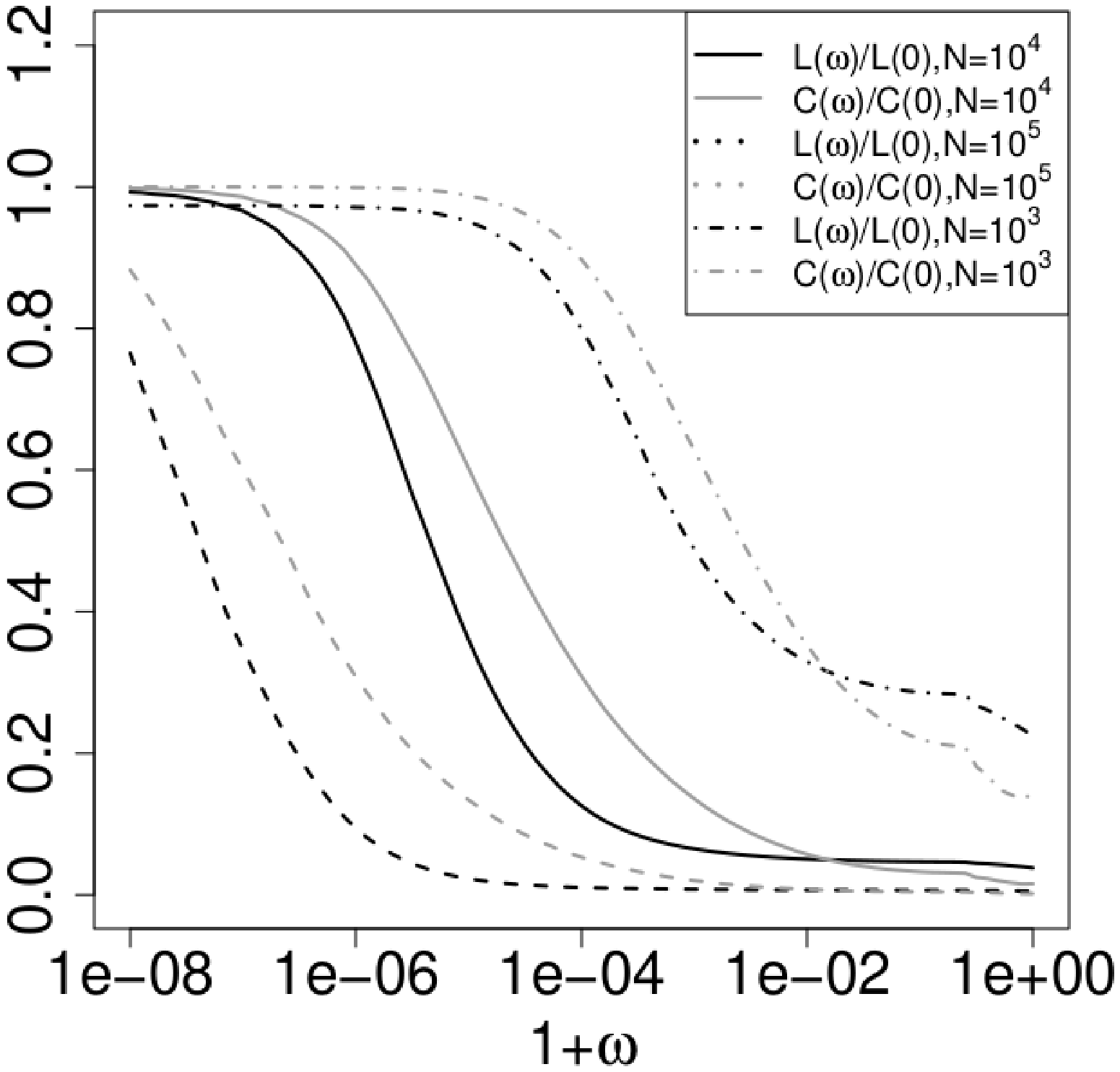}
 \hspace{1.6cm} (b)
\end{center}
\end{minipage}
 \end{tabular}
\caption{
  Average distance $L(\omega)$ and cluster coefficient $C(\omega)$ of the networks.
  (a) $N=10^3,r=3$ and we collect $10^3$ sample networks. WS is the Watts--Strogatz model.
  (b)$N=10^2,10^3,10^4$ and we collect $10^3,10^3$ and $10^2$ networks, respectively.}
\label{cluster}
\end{center}
\end{figure}

We study the average distance of the network $\L(\omega)$
and the cluster coefficient $C(\omega)$ in Fig.\ref{cluster}(a).
We sample $10^3$ networks with $10^3$ nodes and $r=3$, and
  estimate the average distance and the cluster coefficients.
The average distance $\L(\omega)$ decreases rapidly as $\omega$ increases
from $\omega=-1$, which corresponds to an extended lattice.
The cluster coefficient $C(\omega)$ decreases more slowly than
$L(\omega)$, and it becomes minimum at $\omega\sim 0$. 
The regions with large cluster coefficients and small average distances are
observed for $\omega\sim -1$ and $\omega>0$.
These properties are observed in several real-world networks \cite{nw}.
We compared the average distance and the cluster coefficient with those
in the Watts--Strogatz (WS) model, which uses a parameter $p$ instead of $\omega$ \cite{WS}.
$p$ is known as the probability of rewiring the extended lattice.
  When $p=0$ and $p=1$, the network becomes the extended lattice and the random network, respectively.
There is a difference between our model and the WS model in the regions $p\sim 0$ and
$\omega\sim -1$ near the extended lattice where the cluster coefficient is large, and
the average distance is small.
The region with this property is wider in the WS model than in our model.
  We also study the system size dependence of the region in Fig.\ref{cluster}(b).
  As the number of nodes $N$ increases from $10^3$ t0 $10^4$,
  the region moves leftward. The result suggests that the region disappears in the limit $N\to\infty$.
  About the WS mode, the curve of the cluster coefficient does not move with $N$ and only the
  curve for the average distance moves leftward. The region becomes wider the the increase of $N$.
Near the random network $p\sim 1$ and $\omega\sim 0$, we can confirm a small cluster
coefficient and a small average distance.

\subsection{III. Quantum statistics and networks}
In this section, we introduce the fitness model for the nodes \cite{BA1,BA2}
to establish the relation between the degree distribution and the quantum statistics.
The fitness is the weight for popularity $l_i$, and 
the popularity of node $i$ is transformed as $\hat{l}_i=e^{-\beta\epsilon_i} l_i$.
Here, $e^{-\beta\epsilon_i}$ is the fitness, $\beta$ is the inverse temperature,
  and $\epsilon_i$ is the associated energy level of node $i$ in the terminology of
  quantum statistical mechanics.
The equation for the expected number of $k_i$ in Eq.(\ref{df}) is modified as

\begin{equation}
\frac{d l_i}{dt}=
\omega r \frac{e^{-\beta\epsilon_i} l_i}{\sum_i e^{-\beta\epsilon_i} l_i}.
\label{def}
\end{equation}

Here, we assume the solution of this equation as
\begin{equation}
l_i=r\left(\frac{t}{t_i}\right)^{f(\epsilon_i)},
\label{sol}
\end{equation}
which satisfies the initial condition, $l_i=r$ at $t=t_i$.
We substitute Eq.(\ref{sol}) into Eq.(\ref{def}) and
obtain

\begin{equation}
f(\epsilon_i)=\omega \frac{rt e^{-\beta\epsilon_i}}{\sum_i e^{-\beta\epsilon_i} l_i},
\label{f}
\end{equation}

\begin{equation}
 \sum_i e^{-\beta\epsilon_i}l_i\sim
 \int e^{-\beta\epsilon} g(\epsilon)d\epsilon 
 \int_1^t l_idt_i
 = \int e^{-\beta\epsilon} g(\epsilon)r\frac{t-t^f}{1-f(\epsilon)}d\epsilon,
 \label{norm}
\end{equation}
where $g(\epsilon)$ is the distribution of the energy level $\epsilon_i$.

In the large-$t$ limit, we calculate Eq.(\ref{norm}) using $f(\epsilon)<1$,
\begin{equation}
  \sum_i e^{-\beta\epsilon_i} l_i\sim
rt\int   
  g(\epsilon) 
 \frac{e^{-\beta\epsilon} }{1-f(\epsilon)}d\epsilon
 =rte^{-\beta\mu},
 \label{Z2}
\end{equation}
where
\begin{equation}
 e^{-\beta\mu}=\int   
  g(\epsilon) 
 \frac{ e^{-\beta\epsilon}}{1-f(\epsilon)}d\epsilon.   
 \label{17}
\end{equation}
$\mu$ becomes the chemical potential in quantum statistics. 
Thus, we can rewrite Eq.(\ref{f}) using Eq.(\ref{Z2}) as
\begin{equation}
    f(\epsilon)=\omega e^{-\beta(\epsilon-\mu)}.
\end{equation}
Then, we use Eq.(\ref{17}) to obtain
\begin{equation}
1=  \int   
g(\epsilon)  \frac{1}{e^{\beta(\epsilon-\mu)}-\omega}d\epsilon=I(\omega, \beta, \mu).
\label{fb}
\end{equation}
Eq.(\ref{fb}) is the normalization.
$1/(e^{\beta(\epsilon-\mu)}-\omega)$ corresponds to the distribution function in quantum statistics.
It is an extension of the Bose and Fermi distributions and is similar to fractional exclusion statistics \cite{Hal,Bi,Ci}.
When $\omega=1(-1)$, it corresponds to the Bose (Fermi) distribution that
connects the Bose distribution with the Fermi distribution.

\subsubsection{A. Boson-type Network: $\omega>0$} 

We rewrite the distribution function in Eq.(\ref{fb}) as follows: 
\begin{equation}
\frac{1}{e^{\beta(\epsilon-\mu)}-\omega}
=
\frac{1}{\omega(e^{\beta(\epsilon-\mu+\delta_{\mu})}-1)}=\frac{1}{\omega(e^{\beta(\epsilon-\mu ')}-1)},
\label{bo}
\end{equation}
where
$\omega=e^{-\beta\delta_{\mu}}$ and $\mu'=\mu-\delta_{\mu}$.
When $\omega=1$, which corresponds to the BA model,
$\delta_{\mu}=0$, and we can obtain the Bose distribution. 

We consider the BEC of the network \cite{BA1,BA2}
under the condition of large $\beta$.
We can rewrite Eq.(\ref{fb}) using Eq.(\ref{bo}) as
\begin{equation}
1=\frac{1}{\omega}\int  
g(\epsilon)  \frac{1}{e^{\beta(\epsilon-\mu')}-1}d\epsilon=\frac{I'(\omega,\beta,\mu')}{\omega}=   I(\omega, \beta, \mu),
\end{equation}
where 
$I'(\omega,\beta,\mu')=\int   
g(\epsilon) /(e^{\beta(\epsilon-\mu')}-1)d\epsilon$.
We can obtain the condition of the BEC in Eq.(\ref{fb}) as
\begin{equation}
I'(\omega,\beta,\mu')=\omega. 
\label{bec}
\end{equation}

$I'(\omega, \beta, \mu')$ increases as $\mu'$. 
If $I'(\omega, \beta, 0)>\omega$, we obtain the solution $\mu'(<0)$ for Eq.(\ref{bec}).
If $I'(\omega, \beta, 0)\leq \omega$, no solution exists.
$\omega-I'(\omega, \beta, \mu')$ of nodes are connected to the maximum fitness node corresponds to the hub in the network.
This corresponds to the ``the winner takes all’’ or BEC of the network.

To confirm this condition, we consider the case where the
distribution of the energy level $\epsilon_i$ is
\begin{equation}
g(\epsilon)=A \epsilon^{\theta},
\end{equation}
where $\theta$ is a parameter for the power distribution, and $A=(\theta+1)/\epsilon_{max}^{\theta+1}$.
$\epsilon_{max}$ is the maximum energy for the fitness model.
 
The condition of the BEC in Eq.(\ref{bec}) is
\begin{equation}
I'(\omega, \beta,0)=\frac{\theta+1}{(\beta \epsilon_{max})^{\theta+1}}
\int_0^{\beta_{max}}dx\frac{x^{\theta}}{e^x-1}<\omega
\end{equation}
In the large-time limit, we can obtain the lower bound of the transition temperature as
\begin{equation}
T_{BEC}>\epsilon_{max}\left(\frac{\omega}{\zeta(\theta+1)\Gamma(\theta+2)}\right)^{1/(\theta+1)},
\label{tbec}
\end{equation}
where $\Gamma(x)$ is the gamma function and $\zeta(x)$ is the
zeta function.
We can observe the BEC when $\theta>0$.

As $\omega$ increases, the lower bound of the transition temperature increases.
In the limit $\omega\rightarrow \infty$, the phase is the BEC phase.
In the limit $\omega\rightarrow 0$, corresponding to the random network limit, the BEC phase disappears.
In summary, $\omega$ is related to the threshold of the BEC phase by Eq.(\ref{tbec}).
As $\omega$ is increased, lower energy levels are occupied.
In Fig.\ref{bf} (a), we show the depiction of the Bose statistics.

\subsubsection{B. Classical network: $\omega=0$ }
When $\omega=0$, which corresponds to the random network case,
the density function becomes $e^{-\beta(\epsilon-\mu)}$. This corresponds to the 
Boltzmann distribution.

\subsubsection{C. Fermion-type network: $-1\leq\omega<0$}
We revise the distribution function in Eq.(\ref{fb}) as 
\begin{equation}
\frac{1}{e^{\beta(\epsilon-\mu)}+|\omega|}
=
\frac{1}{|\omega|(e^{\beta(\epsilon-\mu+\delta_{\mu})}+1)}=\frac{1}{|\omega|(e^{\beta(\epsilon-\mu ')}+1)},
\label{fe}
\end{equation}
where
$|\omega|=e^{-\beta\delta_{\mu}}$ and 
$\mu'=\mu-\delta_{\mu}$.
When $0<|\omega|\leq 1$, $\delta_{\mu}\geq 0$. 

\begin{figure}[h]
\begin{center}
\begin{tabular}{c}
\begin{minipage}{0.5\hsize}
\begin{center}
\includegraphics[clip, width=8cm]{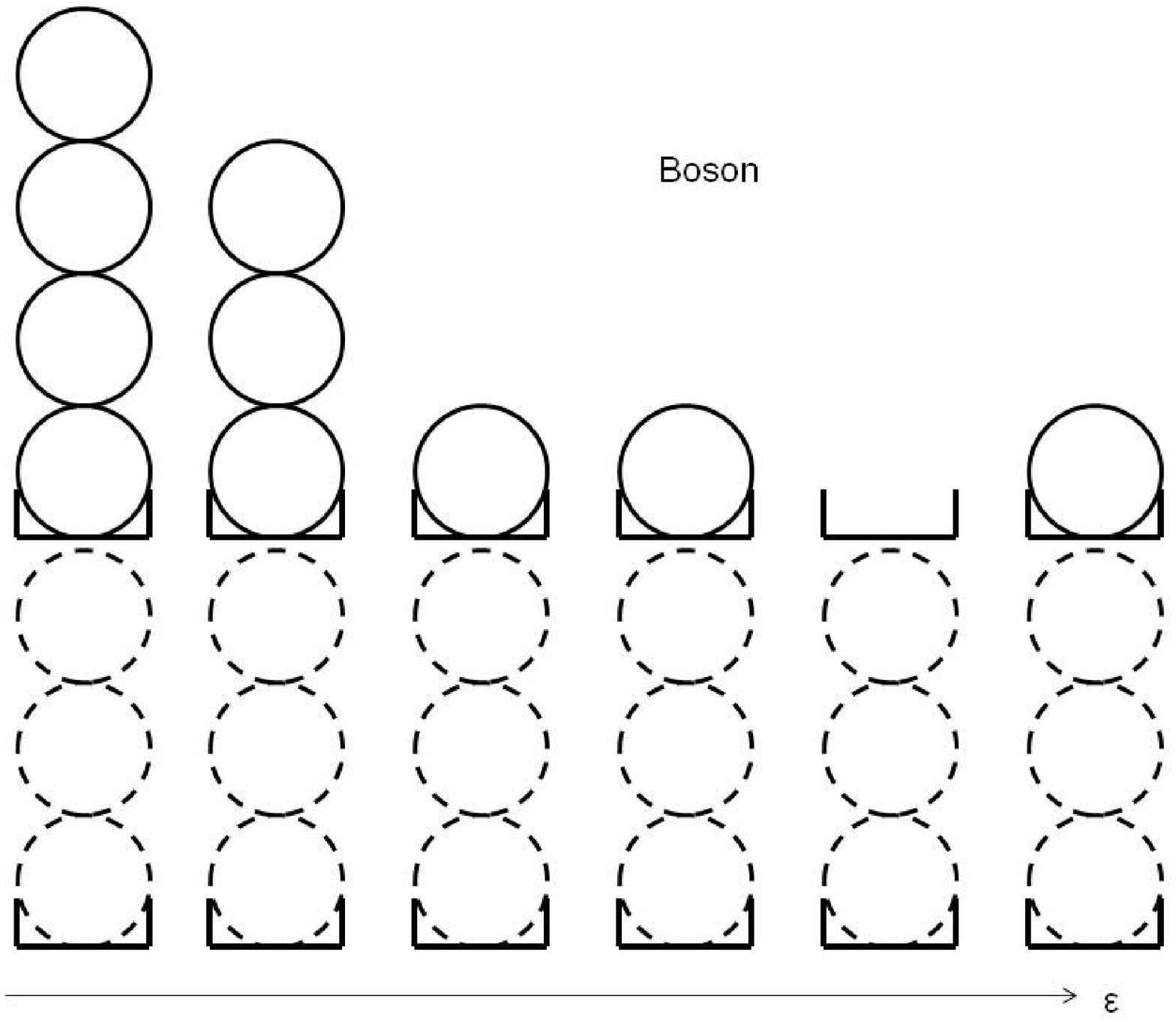}
\hspace{1.6cm} (a)
\end{center}
\end{minipage}
\begin{minipage}{0.5\hsize}
\begin{center}
\includegraphics[clip, width=8cm]{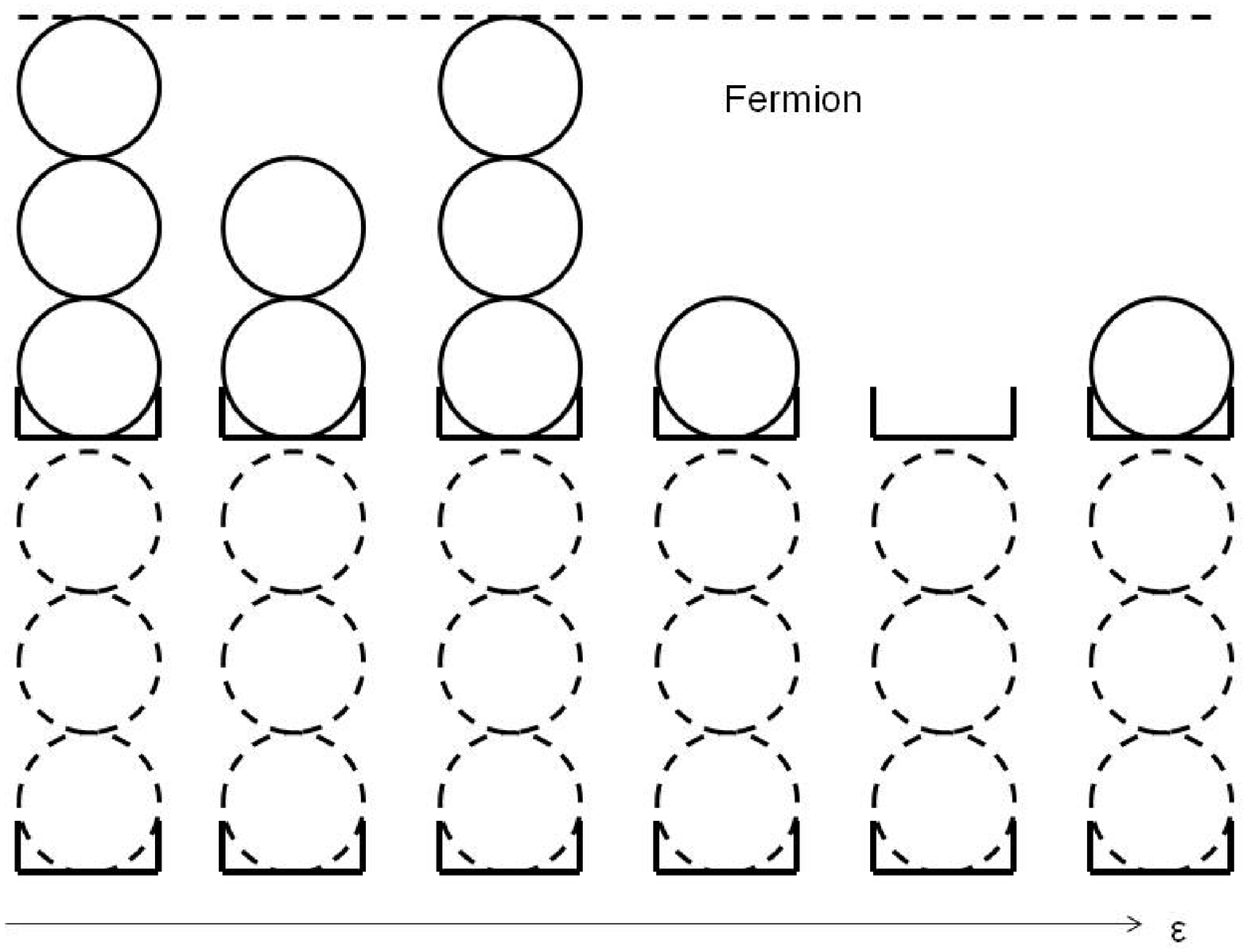}
 \hspace{1.6cm} (b)
\end{center}
\end{minipage}
 \end{tabular}
\caption{(a) Bose statistics. (b) Fermi statistics. The horizontal axis corresponds to $\epsilon$.
  The dotted (real) circles correspond to the incoming (outgoing) arrows.
  In the Fermion case, the occupied number is limited.}
\label{bf}
\end{center}
\end{figure}

We consider the case when $\beta>>1$, i.e., the low-temperature limit.
Eq.(\ref{fe}) becomes the step function at $\epsilon=\mu'$.
When $\epsilon<\mu' (\epsilon>\mu')$, Eq.(\ref{fe}) becomes $1/|\omega| (0)$.
The maximum occupied number, which corresponds to the number of outgoing arrows,
is $-/|\omega|$ per incoming arrow.
$\omega$ corresponds to the change in the chemical potential and
the maximum occupied number in Eq.(\ref{fe}).
On one hand, the lower $\epsilon$ nodes are connected to other $r+\lceil r/|\omega| \rceil$ nodes.
When a node joins the network, the node is connected to $r$ nodes,
and then the node is connected to $\lceil r/|\omega| \rceil$ nodes with the maximum number of links.
On the other hand, the high $\epsilon$ nodes are connected to other $r$ nodes
that are connected only when the node joins.
This is similar to the Fermi degeneracy of the network.
As $\omega<0$ is increased, the maximum occupied number increases,
and the Fermi energy $\mu'$ decreases.
In the limit $\omega\rightarrow 0$, corresponding to the limit of the random
network, there is no maximum occupied number.
In the limit $\omega=-1$, $\delta_{\mu}=0$, we can obtain the Fermi distribution.
The maximum occupied number obtained after the node joins is one per incoming arrow.
In Fig.\ref{bf} (b), we depict the Fermi statistics.

\subsection{VI. Concluding Remarks }

In this article, we discuss several types of networks: 
the random graph, Barab\'{a}si-Albert(BA) model, and lattice networks, using a parameter $\omega$.
$\omega=1,0, and -1$ correspond to the networks above, respectively. The parameter
$\omega$ is related to the preferential attachment of the nodes in the networks.
In our model, we set different weights for the incoming and outgoing arrows.
The parameter $\omega$ is also related to the memory length, which indicates how old a
node is with respect to the new node it connects to.
When $\omega>0$ newly added nodes tend to be connected with older nodes,
and the degree distribution exhibits power-law behavior. $\omega=0$ corresponds to a
random network and long memory with larger power index than the positive $\omega$ case.
Negative $\omega$ corresponds to intermediate memory. The case of $\omega=-1$
corresponds to an extended lattice and short memory.
  
In addition, we discuss the correspondence between quantum statistics
and the degree distribution of the network model. Positive (negative)
$\omega$ corresponds to Bose (Fermi)-like statistics, and we obtain a
distribution that connects the Bose and Fermi statistics.
When $\omega$ is positive, it corresponds to the threshold of BEC.
As $\omega$ decreases, the area of the BEC phase becomes narrower.
When $\omega$ is negative, $r+\lceil r/|\omega| \rceil$ is the maximum occupied number.
At low temperatures, we can observe the Fermi degeneracy of the network.
The lower energy nodes have the maximum number of links, and the higher energy
nodes have no links except for the initial links.
As $\omega$ decreases, the maximum occupied number decreases.

We can observe directed arrow graphs (DAGs) in several fields.
If we consider an arrow from the selected node to the selecting node, our networks become DAGs.
DAGs have recently been used in some cryptocurrencies for scalability.
The scalability is the property of a system that handles the growing confirmation of transactions.
For example, the Bitcoin network is a 1D lattice \cite{bit} without scalability, and
a Tangle (network) of IOTA, which is the network of authentications, is a DAG \cite{iota}.
Tangle has two incoming arrows and two outgoing arrows.
This is an example of a fermionic network with $r=2$ and $\omega=-1$.
For the case of IOTA, the node corresponds to the transaction, and the network
corresponds to the directed links.
Short memory is a necessary property of cryptocurrency networks, because
the confirmations of recent transactions have an incentive.

\end{document}